\documentclass[%
 reprint,
 amsmath,amssymb,
 aps,
]{revtex4-2}

\usepackage{graphicx}
\usepackage{dcolumn}
\usepackage{bm}
\usepackage[utf8]{inputenc} 
\usepackage{braket}
\usepackage{amsmath}
\usepackage{tabularx}

\begin{document}

\preprint{APS/123-QED}

\title{Experimental demonstration of 4-state reference-frame-independent quantum key
distribution over 200km}

\author{Ziran Xie$^{1,2}$}
\author{Zhiyu Tian$^{1,2}$}
\author{Shihai Sun$^{1,2}$}\email{sunshh8@mail.sysu.edu.cn}
\affiliation{%
$^{1}$School of Electronics and Communication Engineering,Sun Yat-sen University, Shenzhen, Guangdong 518107, China\\
$^{2}$Shenzhen Campus of Sun Yat-sen University, No. 66, Gongchang Road, Guangming District, Shenzhen, Guangdong 518107, P.R. China
}%

\begin{abstract}
Reference frame independent quantum key distribution (RFI-QKD) has gained widespread attention due to the unique advantage for practical application, as it circumvents the need for active reference frame alignment within the system. However, in comparison to the standard BB84 protocol, the original 6-state RFI protocol requires a greater number of quantum states to be operated by Alice and Bob, which is an aspect that merits optimization. In this work, we propose a 4-state RFI protocol and illustrate that Alice and Bob each require only four quantum states to perform channel estimation that remains independent of reference frame deviation, which can proficiently reduce the system complexity. Furthermore, through numerical simulations taking the finite-size key effect into consideration, we show that 4-state RFI protocol can achieve a secure key rate and transmission distance on par with the original 6-state RFI protocol. Finally, a experiment over 200 km is inplemented to conducted the feasibility of our scheme. We believe that our protocol can streamline the implementation of RFI-QKD and thereby contribute to the practical advancement of RFI-QKD.

\end{abstract}

\maketitle


\section{INTRODUCTION}
Quantum key distribution(QKD) \cite{bennett1984proceedings} has attracted significant attention and recognition owing to its proven unconditional security in communication \cite{RevModPhys.92.025002}.To enhance the performance of QKD systems under practical conditions (such as secure key rate, maximum transmission distance, and practical security), various improved QKD protocols have been proposed, including the decoy state protocols \cite{PhysRevLett.91.057901,PhysRevA.72.012326,PhysRevLett.94.230504,PhysRevLett.94.230503}, device-independent QKD \cite{Pironio_2009}, measurement device-independent QKD \cite{PhysRevLett.108.130503}, twin-field QKD \cite{TF} and others. Moreover, substantial progress has been made in the practical application of QKD protocol. Although the initial QKD experiment was conducted over a mere distance of 32 cm \cite{bennett1992experimental}, numerous high-rate and long-distance QKD systems have been realized \cite{PhysRevLett.117.190501, earthtoground,PhysRevLett.128.180502,PhysRevLett.130.250801, PhysRevLett.130.250802}. Even intercontinental key distribution spanning up to 4600 km has been achieved utilizing the Micius satellite \cite{earthtoground4600}. Additionally, several enterprises have spearheaded the development of commercial QKD networks and products \cite{PhysRevX.6.011024,10.1049/iet-qtc.2019.0005}.

In practical implementations of QKD systems, the influence of uncontrollable environmental perturbations on physical devices or quantum channels often leads to a inconsistent reference frame between Alice and Bob. Consequently, active compensation are typically introduced to rectify the reference frame deviation between Alice and Bob. Nonetheless, despite their feasibility, these active reference frame compensation inevitably tends to escalate system complexity,which make system more impratical.

In 2010, Laing \textit{et al}. proposed the reference frame independent QKD protocol (RFI-QKD) \cite{PhysRevA.82.012304}. Under the assumption that Z basis is well aligned and the reference frame only variation in the XY plane, an intermediation parameter (called C), which remains invariant to the reference frame variation, is defined to estimate the quantum channel. The XY basis are related by $X_B=cos\beta X_A+sin\beta Y_A$, $Y_B=cos\beta Y_A-sin\beta X_A$, where $\beta$ denotes the relative angle in XY plane.By defining the parameter C as
\begin{equation}
C=\langle X_{A}X_{B} \rangle^{2}+\langle X_{A}Y_{B} \rangle^{2}+\langle Y_{A}X_{B} \rangle^{2}+\langle Y_{A}Y_{B} \rangle^{2},
\end{equation}
it is easy to check that C is independent of $\beta$. Consequently, the maximum information that Eve can intercept can be estimated as $I_{E}=(1-E_{ZZ})h(\frac{1+\mu}{2})+E_{ZZ}h(\frac{1+\nu(\mu)}{2})$,where $h(x)$ is the binary Shannon entropy, and $\mu=min[\frac{\sqrt{C/2}}{1-E_{ZZ}},1]$, $\nu=\sqrt{C/2-(1-{E_{ZZ}}^2 \mu^2)}/E_{ZZ}$. 
By utilizing this method, the necessity for actively reference frame calibration is obviated, thereby reducing system complexity. Hence, the RFI protocol plays a crucial role in some specific scenarios such as satellite to ground and chip-to-chip. There is a rapid development in the practical application of RFI-QKD systems \cite{MDI-RFI,Liu:18,PhysRevApplied.15.064016,Freerunning,9453156}, significantly advancing the practicality of RFI-QKD.

In the original 6-state RFI protocol, Alice and Bob need to handle six quantum states under three different bases to counteract reference frame variations, which increases system complexity compared to the BB84 protocol. To address this limitation of RFI protocol, reducing the number of states required for RFI becomes one of the focuses in advancing RFI protocol implementation. Regarding the research of the fewer states protocol, Wang \textit{et al}. \cite{PhysRevA.92.042319} illustrated that the RFI protocol can function with a mere four quantum states encoded at the transmitter based on the work in \cite{PhysRevA.90.052314,PhysRevA.90.052314}. Moreover, Liu \textit{et al}. \cite{PhysRevApplied.12.034039} proposed a RFI protocol which only needs to send three quantum states $\ket{Z_{0}}$, $\ket{Z_{1}}$, $\ket{X_{0}}$ at the transmitter, and verified the feasibility and effectiveness of the three-state protocol through validation and comparative experiments under 15 km and 55 km fiber channels. Tannous \textit{et al}. \cite{6-4} proposed a 6-4 state protocol where the receiver only needs to select the $Z$ and $X$ bases for measurement and redefined the C value as
\begin{equation}\label{C_64}
C_{64}=\sqrt{\langle X_{A}X_{B} \rangle^{2}+\langle Y_{A}X_{B} \rangle^{2}}
\end{equation}
And through experiment conducted in the entanglement-based scheme, the 6-4 state protocol's feasibility was verified by acquiring the $C_{64}$ value and computing the key rate from the experimental data. Additionally, some studies \cite{Lee:20,PhysRevA.103.022602} also conducted theoretical security analysis and experimental verification on fewer states protocol based on MDI-RFI scheme.

In the aforementioned works, although the number of states manipulated by one side is reduced, the others still needs operate under three bases, which still maintain a higher complexity compared to the standard BB84 protocol. In this paper, we propose a novel RFI protocol where Alice only needs to prepare four states ( $\ket{Z_{0}}$, $\ket{Z_{1}}$, $\ket{X_{0}}$, and $\ket{Y_{0}}$), while Bob only needs to perform measurements in the Z and X bases. Table \ref{tab:example} shows a comparison between our protocol and the aforementioned fewer states protocols.
\begin{table*}[htbp]
\centering
\caption{Requiried quantum state for different RFI QKD protocols}
\label{tab:example}
\begin{tabular}{|c|c|c|}
  \hline
  Protocol & Transmitted states (Alice) & Recived states (Bob) \\
  \hline
Laing \textit{et al.}~\cite{PhysRevA.82.012304} & $\ket{Z_{0}}$, $\ket{Z_{1}}$, $\ket{X_{0}}$, $\ket{X_{1}}$, $\ket{Y_{0}}$, $\ket{Y_{1}}$ & $\ket{Z_{0}}$, $\ket{Z_{1}}$, $\ket{X_{0}}$, $\ket{X_{1}}$, $|Y_0\rangle$, $|Y_1\rangle$ \\
  \hline
  Liu \textit{et al.}~\cite{PhysRevApplied.12.034039} & $\ket{Z_{0}}$, $\ket{Z_{1}}$, $\ket{X_{0}}$ & $\ket{Z_{0}}$, $\ket{Z_{1}}$, $\ket{X_{0}}$, $\ket{X_{1}}$, $\ket{Y_{0}}$, $\ket{Y_{1}}$ \\
  \hline
  Tannous \textit{et al.}~\cite{6-4} & $\ket{Z_{0}}$, $\ket{Z_{1}}$, $\ket{X_{0}}$, $\ket{X_{1}}$, $\ket{Y_{0}}$, $\ket{Y_{1}}$ & $\ket{Z_{0}}$, $\ket{Z_{1}}$, $\ket{X_{0}}$, $\ket{X_{1}}$ \\
  \hline
    Our protocol & $\ket{Z_{0}}$, $\ket{Z_{1}}$, $\ket{X_{0}}$, $\ket{Y_{0}}$ & $\ket{Z_{0}}$, $\ket{Z_{1}}$, $\ket{X_{0}}$, $\ket{X_{1}}$ \\
   \hline
\end{tabular}
\end{table*}
This ensures that the number of quantum states manipulated by both sides is equivalent to that of the standard BB84 protocol. As we only need to prepare $\ket{X_{0}}$ and $\ket{Y_{0}}$ states, the maximum driving voltage required is only $\pi/2$, even lower than the standard BB84 protocol. Additionally, since our protocol only uses the Z and X basis for measurements, we can achieve fully passive measurement at the receiver end without the need for any additional electronic systems. Therefore our protocol  significantly simplify the strutuce of RFI QKD system, also has an advantage in terms of energy consumption. By conducting experiments in the finite-key size scenario, we demonstrate that the protocol exhibits excetional performance over a 200 km fiber channel, further illustrating not only its feasibility but also its remarkable transmission distance in practice. We believe that our proposed 4-state RFI protocol holds significant importance in advancing fewer states RFI-QKD and even in the practical application of RFI-QKD.


\section{PROTOCOL}
In our RFI protocol, Alice randomly sends one of four states ($\ket{Z_{0}}$, $\ket{Z_{1}}$, $\ket{X_{0}}$ and $\ket{Y_{0}}$), and Bob randomly measures the received quantum state with one of two bases (Z and X). The quantum states measured in Z basis are used to generate the secret key, and others are used to estimate Eve's information. Since in our 4-state protocol Bob also only selects the Z and X bases for measurement, we can adopt the definition of the $C_{64}$ value in Eq. (\ref{C_64}) as the intermediate parameter $C_{44}$ for our protocol. However, in the 4-state RFI protocol, where Alice only prepares $\ket{X_{0}}$ and $\ket{Y_{0}}$, as observed in Eq. (\ref{C_64}), it seems that we cannot obtain all the necessary information for the X and Y bases directly from experimental data. Therefore, an alternative method is required for channel estimation. One intuitive approach is to optimize the statistical data measurable in experiment, but this method presents certain challenges. Firstly, the optimization algorithm might exacerbate hardware resource requirements, and secondly, the lower bounds obtained from optimization may not be tight. Below we will illustrate how to obtain the $C_{44}$ value in an analytical form through the Pauli algebra, thereby facilitating the implementation of our 4-state RFI protocol.

In the protocol, without loss of generality, we can assume that Alice's three bases are unbiased with respect to the standard Pauli bases. Specifically, from the entanglement version of the protocol, Alice and Bob share an ideal Bell state $\ket{\psi^{+}}$, and measure their qubits separately. According to this assumption, the measurement operator chosen by Alice can be written with respect to the Z basis in:
\begin{equation}
\begin{gathered}
Z_{0}^A=
\begin{bmatrix}
1 & 0 \\ 0& 0
\end{bmatrix},
Z_{1}^A=
\begin{bmatrix}
0 & 0 \\ 0& 1
\end{bmatrix}\\
X_{0}^A=\frac{1}{2}
\begin{bmatrix}
1 & 1 \\ 1& 1
\end{bmatrix},
Y_{0}^A=\frac{1}{2}
\begin{bmatrix}
1 & -i \\ i& 1
\end{bmatrix}
\end{gathered}
\end{equation}

Therefore, we can obtain:
\begin{equation}\label{XAYA}
\begin{gathered}
X_A=2X_{0}^A-Z_{0}^A-Z_{1}^A\\
Y_A=2Y_{0}^A-Z_{0}^A-Z_{1}^A
\end{gathered}
\end{equation}
Then, for the $C_{44}$ with same definition as Eq. (\ref{C_64}) we have
\begin{equation}\label{C}
C_{44}^2=\langle X_{A}X_{B} \rangle^{2}+\langle Y_{A}X_{B} \rangle^{2}=C_1^2+C_2^2
\end{equation}
According to Eq.~\ref{XAYA}, we have
\begin{equation}\label{C1C2}
\begin{aligned}
C_ 1&=\langle (2X_{0}^A-Z_{0}^A-Z_{1}^A)X_B \rangle \\
&=\langle 2X_{0}^AX_B-Z_{0}^AX_B-Z_{1}^AX_B \rangle\\
C_ 2&=\langle (2Y_{0}^A-Z_{0}^A-Z_{1}^A)X_B \rangle \\
&=\langle 2Y_{0}^AX_B-Z_{0}^AX_B-Z_{0}^AX_B \rangle\\
\end{aligned}
\end{equation}
Therefore, $C_{44}$ can be analytical computed directly using measurable data. Note that the information from the Z basis can also be used to estimate the channel, thus enhancing the protocol's robustness against the statistical fluctuations because of finite-key size. Finally, according to \cite{6-4}, the information leaked to Eve during the communication can be estimated by:
\begin{equation}\label{Ie}
I_{E}=h(\frac{1-C_{44}}{2})
\end{equation}

\section{ESTIMATION OF THE SECRET-KEY RATE}

In most practical QKD systems, a weak coherent light source with random phases is commonly employed, which is vulnerable to photon-number-splitting (PNS) attacks. Therefore, the decoy-state method \cite{PhysRevLett.91.057901,PhysRevLett.94.230503,PhysRevA.72.012326,PhysRevLett.94.230504} is often adopted. Here, we employ the standard “weak+vacuum” decoy-state method, where the intensity $k \in \{\mu, \nu, \omega\}$, subject to the conditions $\mu > \nu + \omega$ and $0 \leq \omega \leq \nu$. Furthermore, due to the finite number of transmitted pulses, the impact of statistical fluctuations must be taken into consideration. Hence in  non-asymptotic case, the secure key length against general attacks is given by \cite{Sheridan_2010,Zhang:17}:
\begin{equation}
\begin{aligned}
l&=\underline{s}_{ZZ}^0+\underline{s}_{ZZ}^1(1-I_E)-n_{zz}fh(E_{ZZ})\\
&-log_2 \frac{2}{\epsilon_{EC}}-2log_2 \frac{2}{\epsilon_{PA}}\\
&-7n_{ZZ}\sqrt{\frac{log_{2}\frac{2}{\overline{\epsilon}}}{n_{ZZ}}}-30log_{2}(N+1)
\end{aligned}
\end{equation}
Here $n_{ZZ}$ and $E_{ZZ}$ represents the number of click events and the quantum bit error rate respectively where Alice(Bob) prepares(measures) quantum states in the Z basis. $\underline{s}_{ZZ}^0$ and $\underline{s}_{ZZ}^1$ denotes the number of vacuum events and single photon events respectively, which should be estimated through the decoy state method. $I_E$ denotes the upper bound of information leaked to Eve, $f$ denotes the efficiency of error correction, $\epsilon_{EC}$($\epsilon_{PA}$) denotes the probability that error correction (privacy amplification) fails, $\overline{\epsilon}$ denotes the accuracy of smooth min-entropy estimation, and \textit{N} denotes the length of the quantum state sent by Alice.

According to the method proposed in \cite{PhysRevA.89.022307}, the counts of vacuum events and single photon events in the Z basis can be given by:
\begin{equation}
\underline{s}_{ZZ}^0=\frac{\tau_0}{\nu-\omega}(\frac{\nu e^{\omega} \underline{n}_{ZZ,\omega}^{*}}{p_\omega}-\frac{\omega e^{\nu} \overline{n}_{ZZ,\nu}^{*}}{p_\nu})
\end{equation}
\begin{equation}
\begin{aligned}
\underline{s}_{ZZ}^1&=\frac{\mu \tau_1}{\mu(\nu-\omega)-(\nu^2-\omega^2)}[\frac{e^\nu \underline{n}_{ZZ,\nu}^*}{p_\nu}\\
&-\frac{e^\omega \overline{n}_{ZZ,\omega}^*}{p_\omega}+\frac{\nu^2-\omega^2}{\mu^2}(\frac{\underline{s}_{ZZ}^0}{\tau_0}-\frac{e^\mu \overline{n}_{ZZ,\mu}^*}{p_\mu})]
\end{aligned}
\end{equation}
Here $\tau_n = \sum_k e^{-k}k_np_k/n!$ represents the probability of Alice sending n-photon pulses, and $\underline{n}_{ZZ,k}^*$ ($\overline{n}_{ZZ,k}^*$) denotes the lower (upper) bound of the average count of $ZZ$ events when Alice selects the source of intensity $k$. Here, we adopt the method proposed in \cite{Yin}. For a given set of Bernoulli variables and observed value $x$, the statistical fluctuation range of the mean $x^*$ can be given by
\begin{equation}
x-\delta_L(\epsilon,x)\leq x^* \leq x-\delta_U(\epsilon,x)
\end{equation}
where $\delta_L(\epsilon,x)=\frac{\beta}{2}+\sqrt{2\beta x+\frac{\beta^2}{4}}$, $\delta_U(\epsilon,x)=\beta+\sqrt{2\beta x+\beta^2}$, $\beta=ln1/\epsilon$. $\epsilon$ represents the probability of the mean value exceeding this range. Thus we have$\underline{n}_{ZZ,k}=n_{ZZ,k}-\delta_L(\overline{\epsilon},n_{ZZ,k})$, $\overline{n}_{ZZ,k}=n_{ZZ,k}-\delta_U(\overline{\epsilon},n_{ZZ,k})$

Below we will illustrate how to estimate the lower bound of $C_{44}$. Firstly, we need to estimate the upper and lower bounds of $C_1$ and $C_2$,denoted as $\overline{C}_{1(2)}$, $\underline{C}_{1(2)}$. Taking $C_1$ as an example, in the non-asymptotic case, according to Equation (\ref{C1C2}), we have:
\begin{equation}
\begin{aligned}
C_1&=\langle (2X_{A}-Z_{A}-Z_{A})X_B \rangle \\
&=(1-2e_{X_0X})-\frac{1}{2}(1-2e_{Z_0X})-\frac{1}{2}(1-2e_{Z_1X}) \\
&=e_{Z_0X}+e_{Z_1X}-2e_{X_0X}
\end{aligned}
\end{equation}
where $e_{\alpha_i \beta}$ represents the single-photon bit error rate when Alice sends $\ket{\alpha_i}$ and Bob measrues in $\beta$ basis. Additionally, let $\overline{e}_{\alpha_i \beta}$($\underline{e}_{\alpha_i \beta}$) denotes the upper(lower) bound of $e_{\alpha_i \beta}$. It is evident that $\overline{C}_{1}$, $\underline{C}_{1}$ satisfy following relation with $\overline{e}_{\alpha_i \beta}$, $\underline{e}_{\alpha_i \beta}$
\begin{equation}
\begin{gathered}
\overline{C}_{1}=\overline{e}_{Z_0X}+\overline{e}_{Z_1X}-2\underline{e}_{X_0X}\\
\underline{C}_{1}=\underline{e}_{Z_0X}+\underline{e}_{Z_1X}-2\overline{e}_{Y_0X}
\end{gathered}
\end{equation}
Similarly, the upper and lower bounds of $C_2$ also satisfy the aforementioned relations with the corresponding $e_{\alpha_i \beta}$. 

Below we will discuss the estimation of the upper and lower bounds of $e_{\alpha_i \beta}$. Notice that $e_{\alpha_i \beta}$ can be expressed by $t_{\alpha_i \beta}/{s}_{\alpha_i \beta}^1$, where $t_{\alpha_i \beta}$ represents the error count of single-photon for the corresponding event, ${{s}_{\alpha_i \beta}^1}$ represents the total count of single-photon for the corresponding event. Thus, the upper and lower bounds of the single-photon error rate can be written as
\begin{equation}
\begin{gathered}
 \underline{e}_{\alpha_i \beta}=\underline{t}_{\alpha_i \beta}/\overline{s}_{\alpha_i \beta}^1 \\\overline{e}_{\alpha_i \beta}=\overline{t}_{\alpha_i \beta}/\underline{s}_{\alpha_i \beta}^1
\end{gathered}
\end{equation}
According to the method mentioned in \cite{PhysRevA.89.022307}, we have:
\begin{equation}
\begin{gathered}
\underline{t}_{\alpha_i \beta}=\frac{\tau_1}{\nu-\omega}(\frac{e^{\nu}\underline{m}_{\alpha_i \beta,\nu}^*}{p_\nu}-\frac{e^{\omega}\overline{m}_{\alpha_i \beta,\omega}^*}{p_\omega})\\
\overline{t}_{\alpha_i \beta}=\frac{\tau_1}{\nu-\omega}(\frac{e^{\nu}\overline{m}_{\alpha_i \beta,\nu}^*}{p_\nu}-\frac{e^{\omega}\underline{m}_{\alpha_i \beta,\omega}^*}{p_\omega})
\end{gathered}
\end{equation}

\begin{equation}
\begin{aligned}
\underline{s}_{\alpha_i \beta}^1&=\frac{\mu \tau_1}{\mu(\nu-\omega)-(\nu^2-\omega^2)}[\frac{e^\nu \underline{n}_{\alpha_i \beta,\nu}^*}{p_\nu}\\
&-\frac{e^\omega \overline{n}_{\alpha_i \beta,\omega}^*}{p_\omega}+\frac{\nu^2-\omega^2}{\mu^2}(\frac{\underline{s}_{\alpha_i \beta}^0}{\tau_0}-\frac{e^\mu \overline{n}_{\alpha_i \beta,\mu}^*}{p_\mu})]\\
\overline{s}_{\alpha_i \beta}^1&=\frac{\mu \tau_1}{\mu(\nu-\omega)-(\nu^2-\omega^2)}[\frac{e^\nu \overline{n}_{\alpha_i \beta,\nu}^*}{p_\nu}\\
&-\frac{e^\omega \underline{n}_{\alpha_i \beta,\omega}^*}{p_\omega}+\frac{\nu^2-\omega^2}{\mu^2}(\frac{\overline{s}_{\alpha_i \beta}^0}{\tau_0}-\frac{e^\mu \underline{n}_{\alpha_i \beta,\mu}^*}{p_\mu})]
\end{aligned}
\end{equation}

\begin{equation}
\begin{gathered}
\underline{s}_{\alpha_i \beta}^0=\frac{\tau_0}{\nu-\omega}(\frac{\nu e^{\omega} \underline{n}_{\alpha_i \beta,\omega}^{*}}{p_\omega}-\frac{\omega e^{\nu} \overline{n}_{\alpha_i \beta,\nu}^{*}}{p_\nu})\\
\overline{s}_{\alpha_i \beta}^0=\frac{\tau_0}{\nu-\omega}(\frac{\nu e^{\omega} \overline{n}_{\alpha_i \beta,\omega}^{*}}{p_\omega}-\frac{\omega e^{\nu} \underline{n}_{\alpha_i \beta,\nu}^{*}}{p_\nu})
\end{gathered}
\end{equation}
where $\underline{n}_{\alpha_i \beta,k}^{*}$($\overline{n}_{\alpha_i \beta,k}^{*}$) represents the lower (upper) bound of the mean count of events when Alice selects a source of intensity $k$ and sends $\ket{\alpha_i}$ while Bob measures in the $\beta$ basis. These bounds can be obtained by considering the observed counts for the corresponding events: $\underline{n}_{\alpha_i \beta,k}^{*}={n}_{\alpha_i \beta,k}-\delta^L(\overline{\epsilon},n_{\alpha_i \beta,k})$,$\overline{n}_{\alpha_i \beta,k}^{*}={n}_{\alpha_i \beta,k}+\delta^U(\overline{\epsilon},n_{\alpha_i \beta,k})$.

Finally, we can move to compute the lower bound of $C_{44}$. From Eq. \ref{C_64} we have 
\begin{equation}
C_{44}^L=\sqrt{{\vert C_1 \vert}^2_L+{\vert C_2 \vert}^2_L}
\end{equation}
Where $C_{44}^L$, ${\vert C_1 \vert}_L$, ${\vert C_2 \vert}_L$ represent the lower bounds of $C_44$,$C_1$ and $C_2$, respectively. It is evident that
\begin{equation}
{\vert C_{1(2)} \vert}_L=
\begin{cases}
\overline{C}_{1(2)} & if \ \overline{C}_{1(2)},\underline{C}_{1(2)}<0 \\
\underline{C}_{1(2)} & if \ \overline{C}_{1(2)},\underline{C}_{1(2)}>0 \\
0 & else 
\end{cases}
\end{equation}
Thus after computing the upper and lower bounds of $C_1$ and $C_2$,  we can now finalize the estimation of the lower bound of $C_{44}$ while considering statistical fluctuations.

\begin{figure}[htb]
\includegraphics[width=1\linewidth]{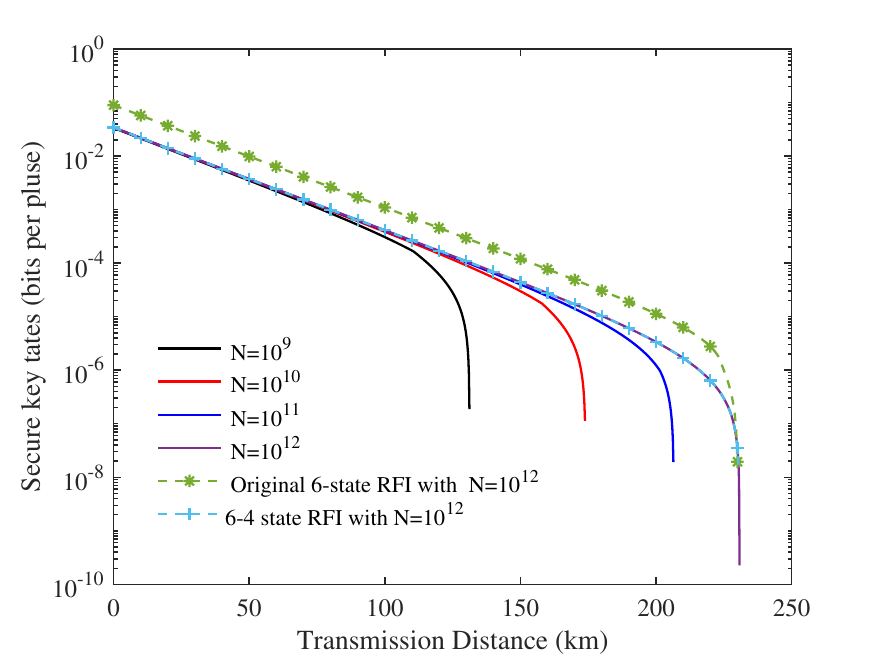}
\caption{\label{fig2}  The performance of our 4-state RFI protocol in terms of the key rate. The solid lines represent the key rate of the 4-state RFI protocol at different data length. The dashed lines depict the key rate curves of the original 6-state RFI protocol \cite{PhysRevA.82.012304} and the 6-4 state RFI protocol \cite{6-4} with the same parameters. Here, we assume that the probabilities of Alice sending states $\ket{X_0}$, $\ket{X_1}$, $\ket{Y_0}$, and $\ket{Y_1}$ are all $(1-p_Z)/4$. In the 6-6 state protocol,we set the probabilities for Bob of selecting the Z, X, and Y basis are 0.5, 0.25, and 0.25, respectively.}
\end{figure}

For evaluation, we conducted numerical simulations for the final key rate $r=l/N$ under finite-key size. The simulation parameters were set based on our experimental system and fully listed in Table\ref{table1}. We set $p_{X_0}=p_{Y_0}=(1-p_Z)/2$,where $p_{X_0}$ and $p_{Y_0}$ are the probabilities of Alice preparing the states $\ket{X_0}$ and $\ket{Y_0}$ respectively and $p_Z$ is the probability of Alice select Z basis.  Fig.\ref{fig2} illustrates the performance of the 4-state RFI protocol for different data length. For comparison, we simulated the key rates of the 6-state RFI protocol. It is observed that except for a slight decrease in key rate, there is almost no degradation in performance. We also conducted the simulation of the 6-4 state RFI protocol and exhibit that the performance of 4-state RFI protocol is nearly identical to the 6-4 state RFI protocol.

\begin{table}
\caption{\label{table1}
Parameters in simulations, which derived from our experimental set up. $e_0$ is the intrinsic error rate of the system, $\alpha$ is the loss coefficient of fiber, $\eta_Z$ and $\eta_{XY}$ are the loss of Z path and XY path in receiver, respectively. $e_d$ is the dark count rate, $\eta_d$ is the efficiency of detector, $f$ is the efficiency of error corection, $\overline{\epsilon}$ is the accuracy of smooth min-entropy estimation, and we set that $\overline{\epsilon}=\epsilon_{EC}=\epsilon_{PA}$.}
\begin{ruledtabular}
\begin{tabular}{cccccccc}
\textrm{$e_0$}&\textrm{$\alpha$}&\textrm{$\eta_Z$}&\textrm{$\eta_{XY}$}&\textrm{$e_d$}&\textrm{$\eta_d$}&\textrm{$f$}&\textrm{$\overline{\epsilon}$}\\
\colrule
$1\%$ & 0.19 dB/km & 4 dB & 9 dB & $1.3\times 10^{-7}$ & $60\%$ & 1.1 & $10^{-10}$\\

\end{tabular}
\end{ruledtabular}
\end{table}

\section{EXPERIMENTAL SETUP AND RESULT}
\begin{figure}[htb]
\includegraphics[width=1\linewidth]{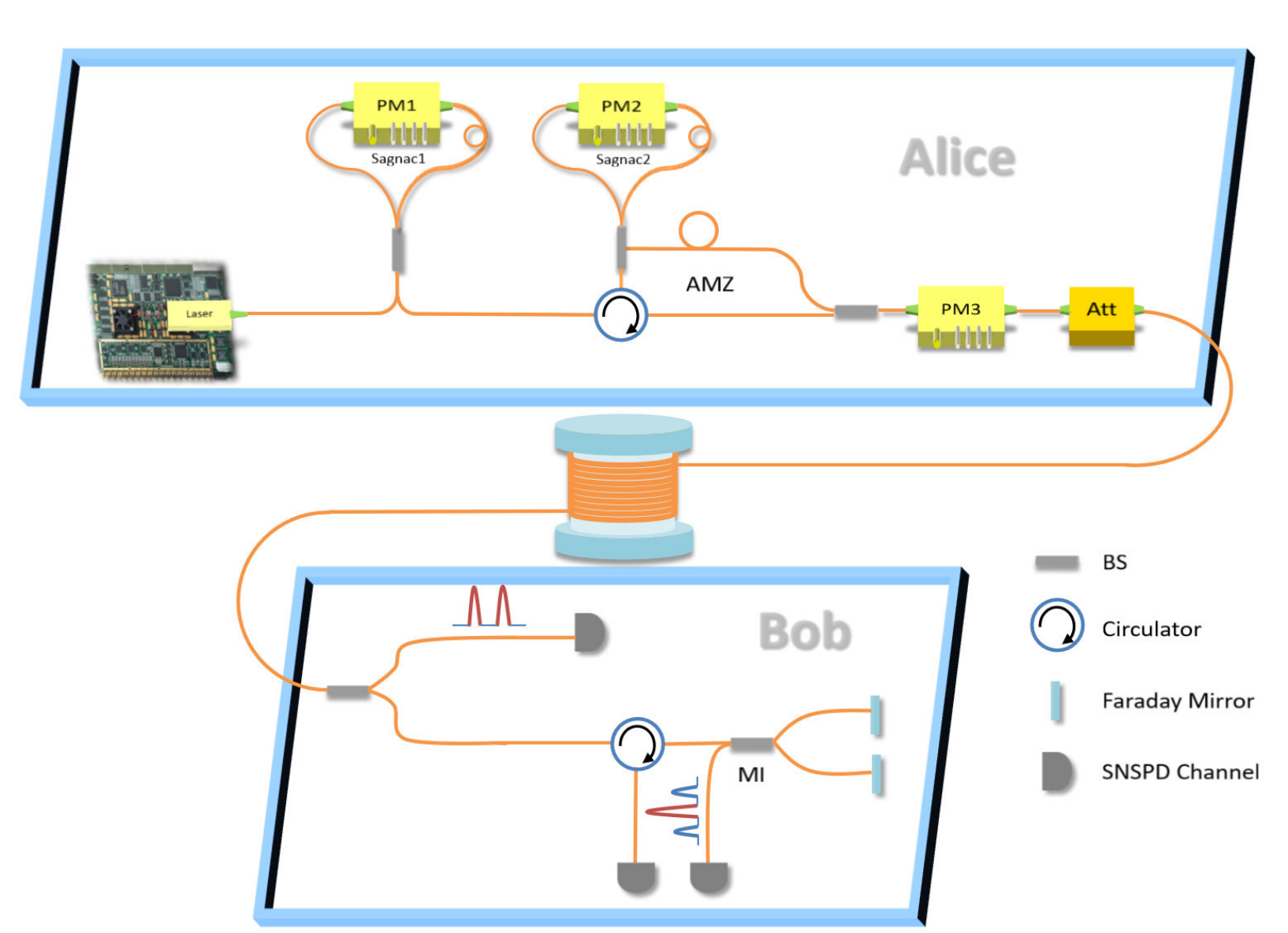}
\caption{\label{fig3} Experiment set up diagram for our experiment system. BS:Beam Splitter. PM:Phase Modulation. Att:Optical Attenuator. SNSPD:Superconducting Nanowire Single Photon Dectector.}
\end{figure}

The experimental setup is depicted in Fig. \ref{fig3} based on time-bin coding. 
At the transmitter, we utilized Sagnac interferometer to generate different decoy states. By adjusting the voltage applied to PM1, the intensity of the optical pulses at the output of Sagnac1 can be manipulated. After Sagnac1, the Sagnac2 in conjunction with an asymmetric Mach-Zehnder interferometer (AMZ) with a relative delay of 3 ns were utilized to encode quantum states. By varying the voltage on PM2 between phase 0 or $\pi$, the optical pulses were directed through the short or long path of AMZ, generating the states $\ket{Z_0}$ or $\ket{Z_1}$. When phase $\pi/2$ was modulated on PM2, the optical pulses with same intensity of two path was generated. Then, by modulating one of the optical pulse arms using PM3 to induce a relative phase difference between the two optical pulses, different quantum states in the XY bases can be prepared. Specifically, Alice only needs to use voltage inputs of $\{ 0,\pi/2    \}$ on PM3 to prepare states $\{ \ket{X_0},\ket{Y_0}    \}$ in XY bases. It is worth noting that in our protocol, since Alice only selected states $\ket{X_0}$ and $\ket{Y_0}$ in the XY basis, the highest voltage loaded on PM3 was only $\pi/2$, which is lower than the initial RFI protocol's $3\pi/2$ voltage and even the BB84 protocol's $\pi$ voltage. Therefore, our protocol has an advantage in terms of energy consumption, especially in high-repetition-rate systems, where this advantage becomes more pronounced.

At the reciever, Bob passively selects either the Z basis or the X basis for demodulation using a beam splitter (BS). For Z basis, the optical pulse is directly input into the SNSPD who works in the free running model. Thus, the quantum state $\ket{Z_0}$ and $\ket{Z_1}$ can be distinguished with the time stamp. This simple structure could remove an additional BS and reduce the loss of Bob's system. For X basis, Bob decoded the quantum states by utilizing a Faraday-Michelson interferometer (MI) equipped with relative delays corresponding to the AMZ. Since our protocol does not necessitate the measurement in Y basis, no PM is required at the reciver thereby reducing the complexity of system.

We integrate the electronic module along with the laser diode onto a homemade FPGA. The laser emitting optical pulses at a central wavelength of 1550 nm with the width of 50 ps and the repetition rate of 100 MHz. The data processing of the system is predominantly managed by a PC equipped with a Time-to-Digital Converter (TDC). Its responsibilities include synchronizing, selecting detector time windows, raw key sifting, and performing statistical analysis of detection results. And an external 10 MHz square wave signal serves as the reference clock between the FPGA and TDC. To demonstrate the performance of our protocol, we conduct a long-distance experiments over 200 km fiber channel. The detailed experimental results is presented in Table \ref{table2}.
\begin{table*}[htbp]
\caption{\label{table2}
The experimental results.$r_L$ represents the lower bound of the secure key rate. $\widehat{C}_L$ represents the mean lower bound of the estimated $C_{44}$ value per data group. $s_{ZZ}^1$ represents the lower bound of the single photon count estimated dor the Z basis. $E_{ZZ}$ represents the QBER for the Z basis. 
}
\begin{ruledtabular}
\begin{tabular}{cccccccccccc}
\textrm{Distance}&\textrm{$r_L$}&\textrm{$\widehat{C}_L$}&\textrm{$s_{ZZ}^1$}&\textrm{$E_{ZZ}$}&\textrm{$p_Z$}&\textrm{$\mu$}&\textrm{$\nu$}&\textrm{$\omega$}&\textrm{$p_\mu$}&\textrm{$p_\nu$}&\textrm{$p_\omega$}\\
\colrule
200km & $3.04\times 10^{-6}$ & 0.6503 & $3.3\times 10^7$ & $0.77\%$ & 0.77 & 0.55 & 0.28& 0& 0.54& 0.36& 0.10\\

\end{tabular}
\end{ruledtabular}
\end{table*}

Due to the influence of statistical fluctuations, a large code length is required to achieve acceptable key rates in the case of long distances. Consequently, the system needs to operate for extended periodsz of time. Prolonged operation implies that the deviation angle $\beta$ may undergo significant drift, and lead to the failure of the RFI protocol \cite{PhysRevA.94.062330}. Therefore,  we adopt the method proposed in \cite{Freerunning} during the experiments.  In practical RFI systems, the QBER under X and Y bases varies approximately sinusoidally with $\beta$ \cite{Freerunning}. Based on this mapping relationship, a classification parameter $\rho \in [0,2\pi]$ can be assigned to the data for each time interval (e.g., 1 second). This parameter is defined as:
\begin{equation}
\rho=
\begin{cases}
arccos(\frac{2}{\eta \mu}ln\frac{H(\widehat{E}_{\mu}^{XX})}{2\widehat{E}_{\mu}^{XX}}-1) & E_{\mu}^{XY}<0.5 \\
2\pi-arccos(\frac{2}{\eta \mu}ln\frac{H(\widehat{E}_{\mu}^{XX})}{2\widehat{E}_{\mu}^{XX}}-1) & E_{\mu}^{XX} \geq 0.5 
\end{cases}
\end{equation}
where
\begin{equation}
\begin{aligned}
H&(\widehat{E}_{\mu}^{XX})=(1-e_d)(2\widehat{E}_{\mu}^{XX}-1)\\
&+\sqrt{4e^{\eta \mu}\widehat{E}_{\mu}^{XY}(1-\widehat{E}_{\mu}^{XX})+{(1-e_d)}^2{(1-2\widehat{E}_{\mu}^{XX})}^2}
\end{aligned}
\end{equation}
\begin{equation}
\widehat{E}_{\mu}^{XX}=\frac{E_{\mu}^{XX}-e_0}{1-e_0}
\end{equation}
where $E_{\mu}^{XX}$ ($E_{\mu}^{XY}$) represents the QBER when Alice selects the X basis and Bob measures in the X (Y) basis. $\eta$ denotes the channel loss rate, and $\mu$ represents the intensity of the signal state. Then we can uniformly split the classification parameter $\rho$ into M slice therefore classified the data into M datasets. The data belonging to the same interval of $\rho$ are grouped together, and the secure key length is calculated for each group respectively. Finally, the key lengths for all groups are summed to obtain the total secure key length. It is worth noting that $\rho$ does not precisely correspond to the deviation angle of the reference frame for the data group. Rather, it serves to characterize the similarity in the deviation situation of data with similar $\rho$ values. Therefore, the deviation angles within the same data group do not differ significantly, ensuring the effectiveness of RFI.

\begin{figure}[htb]
\includegraphics[width=1\linewidth]{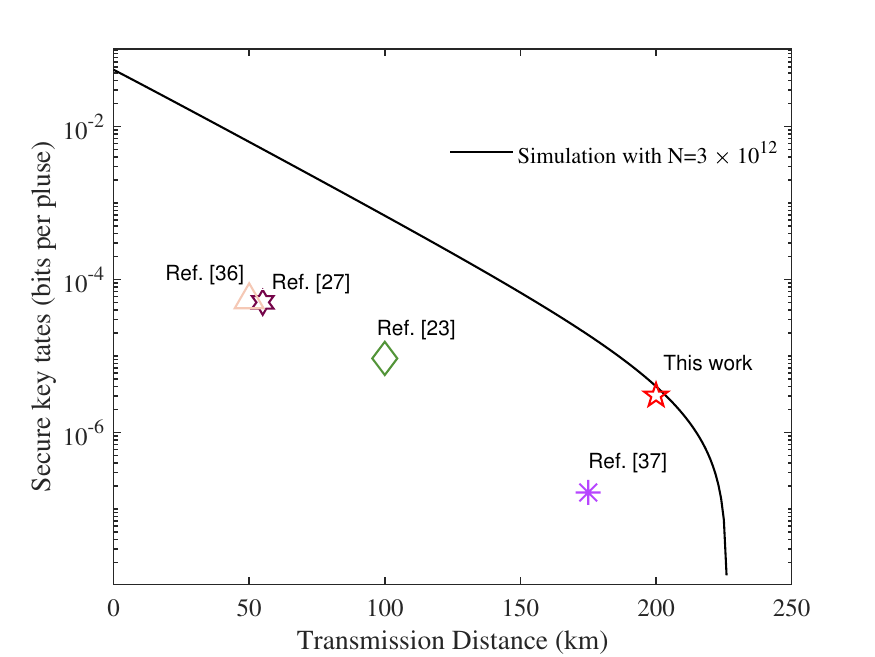}
\caption{\label{fig4} The comparison between experimental results and simulations. The black solid line represents the simulation result.}
\end{figure}

In the experiment, we set the total transmission length to $N=3\times 10^{12}$ and the group number to $M=6$. Fig.\ref{fig4} presents the simulation of the key rate versus distance, along with experimental results and comparisons with recent experiments related to RFI protocols.

\section{CONCLUSION}
The RFI protocol has unique advantages and tremendous potential for practical applications because it eliminates the need for active compensation of the reference frame. Since its inception, researchers have been striving to improve the performance of RFI protocols in practical applications, such as increasing the key rate and distance, enhancing the robustness under finite-key size, and reducing the complexity of system deployment.In this work, we propose an RFI-QKD protocol where Alice and Bob only need to handle four quantum states. Compared to the original RFI protocol, in 4-state RFI protocol, Alice's modulation voltage at the XY basis only needs a maximum of $\pi/2$ voltage, reducing the system's energy consumption. Additionally, the need for PM is completely eliminated at Bob, reducing system complexity and inherent errors arising from imprecise modulation voltage. Subsequently, through numerical simulations considering the decoy-state method and finite code length, we accessed the  performance of 4-state RFI protocol. Finally, we conducted an experiment based on time-bin encoding  over a 200 km fiber channel. The experimental results showed that the secure key rate was consistent with theoretical predictions, demonstrating significant tolerance to the channel loss. Consequently, our work advances the practical application of RFI-QKD systems, enabling RFI systems to be deployed with lower complexity in scenarios where cost control and stability requirements are more stringent.

\section{Acknowledgments}
The authors thank Feihu Xu for very useful discussions. This study was supported by National Natural Science Foundation of China (Grant No. 62171458),  Shenzhen Science and Technology Program (Grant No. JCYJ20220818102014029).

\nocite{*}

\bibliography{apssamp2_En}

\end{document}